\documentclass[letterpaper,twocolumn,shortnote]{jpsj3}
\usepackage{txfonts}

\usepackage{xcolor}

\usepackage[final]{changes}

\usepackage{txfonts}
\usepackage{bm}

\definechangesauthor[name = Dai, color=red]{DA}
\definechangesauthor[color=green]{JF}
\definechangesauthor[name = Georg, color = magenta]{GK}

\title{de Haas-van Alphen Oscillations for the Field Along $c$-axis in UTe$_2$}

\author{
Dai~Aoki$^1$\thanks{E-mail: aoki@imr.tohoku.ac.jp}, 
Ilya~Sheikin$^2$,
Alix~McCollam$^3$,
Jun~Ishizuka$^4$,
Youichi~Yanase$^5$,
Gerard~Lapertot$^6$,
Jacques~Flouquet$^6$, and
Georg~Knebel$^6$
}

\inst{%
$^1$IMR, Tohoku University, Oarai, Ibaraki 311-1313, Japan\\
$^2$LNCMI-EMFL, CNRS, UGA, 38042 Grenoble, France\\
$^3$HFML-EMFL, Radboud University, Toernooiveld 7, 6525 ED Nijmegen, The Netherlands\\
$^4$Faculty of Engineering, Niigata University, Ikarashi, Niigata 950-2181, Japan\\
$^5$Department of Physics, Graduate School of Science, Kyoto University, Kyoto 606-8502, Japan\\
$^6$Univ. Grenoble Alpes, CEA, Grenoble INP, IRIG, PHELIQS, F-38000 Grenoble, France\\
}

\abst{%
We performed \deleted[id=GK]{the} de Haas-van Alphen (dHvA) experiments \replaced[id=GK]{in}{on} the spin-triplet superconductor UTe$_2$ for \replaced[id=GK]{magnetic}{the} field along \added[id=GK]{the} $c$-axis above $15\,{\rm T}$.
Three fundamental dHvA frequencies, named $\alpha_1$, $\alpha_2$ and $\beta$ corresponding to the cross sections of cylindrical Fermi surfaces (FSs) with \deleted[id=GK]{the} large cyclotron effective masses ($33$--$43\,m_0$) were detected.
No other fundamental dHvA frequencies were detected at high frequency range, suggesting a cylindrical-shaped electron FS without connecting at the $Z$ point of the Brillouin zone.
However, the existence of small pocket FSs associated with extremely heavy masses cannot be \added[id=GK]{fully} excluded.
}

\begin{document}
\maketitle
The heavy-fermion paramagnet UTe$_2$ is one of the hottest materials in condensed matter physics, see recent review.~\cite{Aok22_UTe2_review} 
The superconducting transition occurs at $T_{\rm c}=1.6\mbox{--}2.1\,{\rm K}$.
The highlight is the huge upper critical field, $H_{\rm c2}$, 
highly exceeding the Pauli limit for all field directions 
associated with the field-reentrant behavior for $H\parallel b$-axis.
Another remarkable point is the multiple superconducting phases under pressure and in magnetic fields, detected as a thermodynamic response. 
These results strongly suggest spin-triplet superconductivity in UTe$_2$.
Furthermore, possible topological superconductivity was suggested both theoretically and experimentally.

In order to understand unconventional superconductivity in UTe$_2$, it is important to clarify the electronic \replaced[id=GK]{structure}{states} from a microscopic point of view.
Recently, we reported the first observation of the de Haas-van Alphen (dHvA) effect, \replaced[id=GK]{clarifying}{and clarified} two kinds of cylindrical Fermi surfaces (FSs) associated with large cyclotron effective masses ($32$--$57\,m_0$).~\cite{Aok22_UTe2_dHvA}
However, the dHvA signal for $H\parallel c$-axis was missing, because of the large $H_{\rm c2}$ exceeding our maximum field \added[id=GK]{of} $15\,{\rm T}$.

In this report, we present \deleted[id=GK]{our} results of \deleted[id=GK]{the} dHvA experiments at high fields up to $30\,{\rm T}$ for $H \parallel c$-axis in UTe$_2$.
High-quality single crystals of UTe$_2$ with the residual resistivity ratio, $\mbox{RRR}\sim 400$--$800$ in this batch, were grown using the modified NaCl/KCl-flux method similar to Ref.~\citen{Sak22}.
The dHvA experiments were performed by field-modulation technique in a dilution refrigerator at temperatures down to $ 75\,{\rm mK}$ and at high fields up to $30\,{\rm T}$ at the HFML in Nijmegen.

\begin{figure}[t]
\begin{center}
\includegraphics[width= 1\hsize]{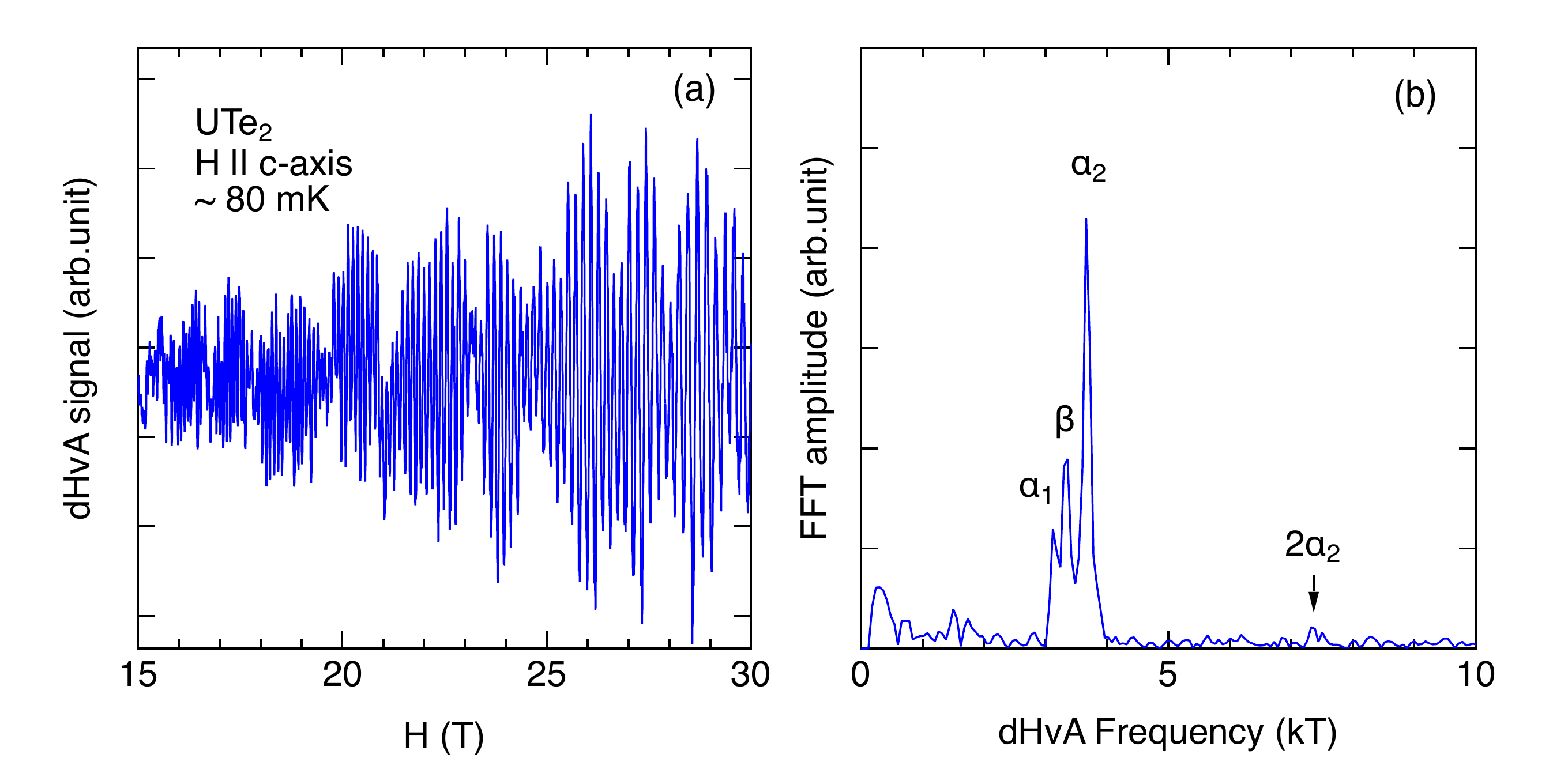}
\end{center}
\caption{(Color online) (a) dHvA oscillations after subtracting non-oscillating background around $80\,{\rm mK}$ and (b) the corresponding FFT spectrum for the field range from $20$ to $30\,{\rm T}$ in UTe$_2$.}
\label{fig:UTe2_Osc_FFT}
\end{figure}

Figure~\ref{fig:UTe2_Osc_FFT} shows the dHvA oscillations and the corresponding FFT spectrum for the field range from $20$ to $30\,{\rm T}$.
Three fundamental dHvA branches $\alpha_2$, $\beta$ and $\alpha_1$ were detected at $3.67$, $3.33$ and $3.14\,{\rm kT}$, respectively,
together with the 2nd harmonic of the branch $\alpha_2$ at $7.35\,{\rm kT}$.
\\
The cyclotron effective masses were determined to be $43\,m_0$, $39\,m_0$ and $33\,m_0$ for branches $\alpha_2$, $\beta$ and $\alpha_1$, respectively,
from the dHvA measurements at different temperatures up to $150\,{\rm mK}$.


\begin{fullfigure}[htbh]
\begin{center}
\includegraphics[width= 0.8\hsize,clip]{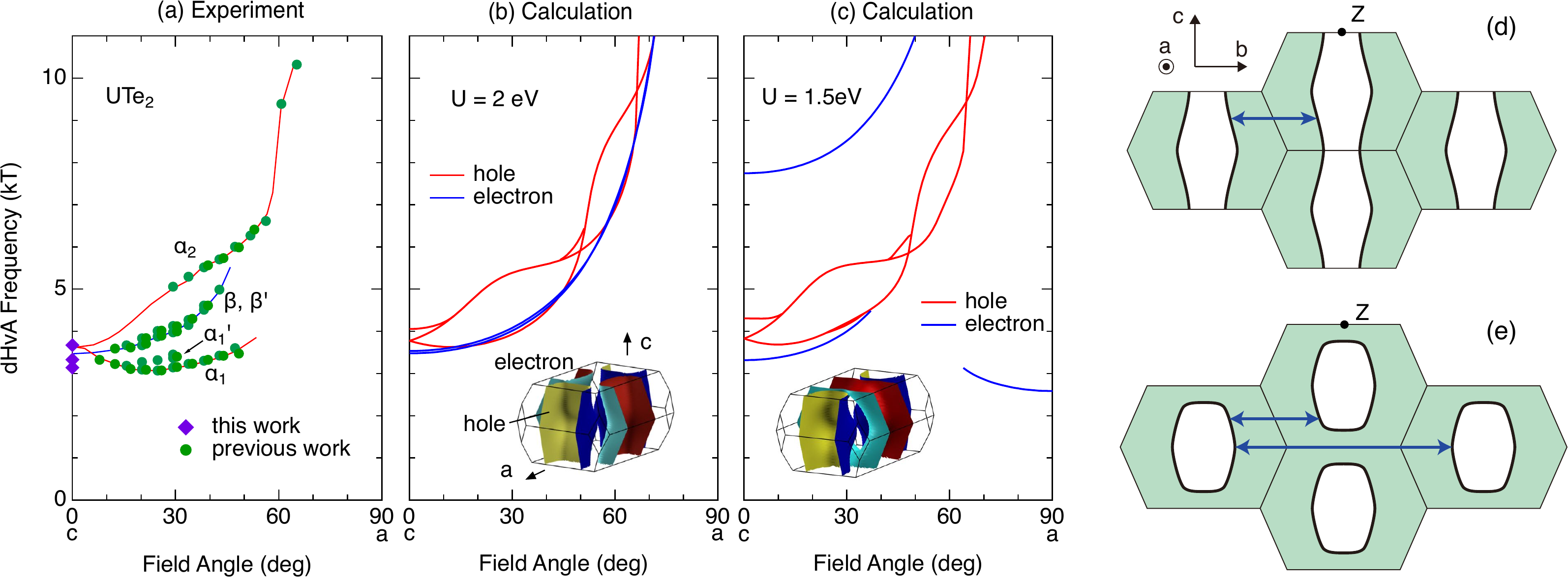}
\end{center}
\caption{(Color online) (a) The detected dHvA frequencies for $H\parallel c$-axis plotted on the previous angular dependence of frequencies in UTe$_2$~\cite{Aok22_UTe2_dHvA}. Panels (b) and (c) show the calculated angular dependence of dHvA frequencies by GGA+$U$ with $U=2\,{\rm eV}$ and $1.5\,{\rm eV}$, respectively. The insets display the corresponding FS. Panels (d) and (e) schematically show the cross sections for electron FS viewed from $a$-axis for $U=2\,{\rm eV}$ and $1.5\,{\rm eV}$, respectively.
Arrows correspond to the diameters of the cyclotron motions for the extremal cross-sectional areas for $H\parallel c$.
}
\label{fig:UTe2_AngDep_FS}
\end{fullfigure}
The obtained dHvA frequencies for $H\parallel c$-axis are added to the angular dependence of the previously reported dHvA frequencies~\cite{Aok22_UTe2_dHvA} in Fig.~\ref{fig:UTe2_AngDep_FS}(a).
The dHvA frequencies for $H \parallel c$-axis agree well with the previous results as an extension to $c$-axis.
Figure~\ref{fig:UTe2_AngDep_FS}(b) shows theoretical angular dependence of the dHvA frequencies calculated by the GGA+$U$ methods with Coulomb repulsion, $U=2\,{\rm eV}$.
The corrugated cylindrical FS for the branch $\alpha$ yields the splitting of the dHvA frequency due to the maximum and minimum cross-sectional areas.
On the other hand, the frequencies for the branch $\beta$ are nearly degenerate even for $H\parallel c$-axis.
This can be understood from a wavy-shaped FS in the Brillouin zone based on the body-centered orthorhombic structure, as shown in Fig.~\ref{fig:UTe2_AngDep_FS}(d).
If the corrugation is moderate, the dHvA frequencies for maximal and minimal cross-sectional areas could be almost degenerate even for $H\parallel c$-axis.
Meanwhile, the FS for the branch $\alpha$ is strongly corrugated, and the frequency splits into two, $\alpha_1$ and $\alpha_2$, for $H\parallel c$-axis \added[id=GK]{in our experiment}.

No \added[id=GK]{higher fundamental} dHvA frequency was observed, while the 2nd harmonic of the branch $\alpha_2$ was detected at $7.35\,{\rm kT}$. 
An important question is whether the FSs of UTe$_2$ consist of only \added[id=GK]{these detected}{} cylindrical FSs. 
In \replaced[id=GK]{the}{our previous} calculation with smaller $U$ ($=1.5\,{\rm eV}$), 
the electron FSs are connected at the $Z$ point, forming a ring-shaped FS, as shown in the inset of Fig.~\ref{fig:UTe2_AngDep_FS}(c).
The corresponding angular dependence of the dHvA frequencies will consist of two sets of \added[id=GK]{frequencies with}{} $1/\cos\theta$-like behavior as a function of field angle $\theta$ from $c$ to $a$-axis. 
The predicted higher frequency is about $8\,{\rm kT}$ for $c$-axis, in which the cyclotron motion is stretched to different Brillouin zones, as shown in Fig.~\ref{fig:UTe2_AngDep_FS}(e).
The calculated band mass for this higher frequency is nearly twice larger than that for the lower one.
Thus, the corresponding dHvA signal could be strongly damped, if it exists.
On the other hand, the low frequency ($\sim 2.5\,{\rm kT}$) close to $H\parallel a$, originating from a ring-shaped orbit, must be easily detected because of the relatively low band mass with a favorable curvature factor.
However, no dHvA signal was detected for $H\parallel  a$ in our dHvA experiments using field-modulation technique either in a superconducting magnet or in a resistive magnet up to $30\,{\rm T}$.

Therefore, next important issue is how to reconcile the two-dimensional FS with the anisotropy of $H_{\rm c2}$.
The initial slope of $H_{\rm c2}$, $H_{\rm{c2}}^\prime$ ($\equiv | dH_{\rm c2}/dT|_{T=T_{\rm c}}$), should be proportional to $1/v_{\rm F}^2$.
Thus, the anisotropy of $H_{\rm{c2}}^\prime$ 
would be explained by the so-called effective mass model~\cite{Aok22_UTe2_review}, in which the topology of the averaged FS associated with the anisotropic effective mass determines the anisotropy of $H_{\rm c2}$.
Cylindrical FS elongated along the $c$-axis suggests  a low $H_{\rm{c2}}^\prime$ for $H \parallel c$-axis and high $H_{\rm{c2}}^\prime$ for $H \parallel a$ and $b$-axes.
In fact, the lowest value of  $H_{\rm{c2}}^\prime$ for $H\parallel c$-axis, ($\sim 7.5\,{\rm T/K}$) and higher values for $b$ and $a$-axis ($20$--$35\,{\rm T/K}$) are reported from the results of specific heat measurements.~\cite{Rosuel23}.
The anisotropy is, however, still small, compared to those expected from two-dimensional FSs.

\added[id=GK]{Furthermore}{}, resistivity also shows rather isotropic behavior in terms of the current direction.~\cite{Eo21}
The resistivity ratio between $J \parallel c$ and $a$ ($b$) is only $\rho_c/\rho_a\sim 2$\quad ($\rho_c/\rho_b\sim 1$) at room temperature.
The value of $\rho_c/\rho_a$ increases up to $10$ at low temperatures, suggestive of electronic structure changes as a function of temperature, but it is still not very large. 

From the present data we cannot rule out the existence of a small FS pocket associated with a very heavy effective mass, which is not detected in the dHvA experiments. 
Two cylindrical FSs yield the total Sommerfeld coefficient, $\gamma \sim 100 {\rm mJ\,K^{-2}mol^{-1}}$, indicating that the main FS are detected. 
If we expect that the $\gamma$-value of about $20\,{\rm mJ\,K^{-2}mol^{-1}}$ is missing,
one can assume the existence of a spherical FS pocket at the Z point, for instance, with a dHvA frequency of $0.2\,{\rm kT}$ and an effective mass $90\,m_0$, 
for which the $\gamma$-value is calculated by $\gamma=k_{\rm B}^2 V/(3\hbar^2) m^\ast k_{\rm F}$.
It is hard to detect such a small FS pocket with heavy mass in the dHvA experiments, especially by the field-modulation technique. 

Recently, quantum oscillations were detected by torque method for the field directions from $c$ to $a$-axis as well as from $c$ to $b$-axis.~\cite{Eat23} 
Theses results confirm our previous conclusions for two cylindrical FS. 
Based on the angular dependence of the quantum oscillation frequencies observed in their experiment and our previous results \cite{Aok22_UTe2_dHvA} they have modeled the cylindrical FS topology with super-elliptical cross-sections with no corrugation for the hole FS. 
The observation of only one dHvA frequency along the $c$-axis, at odds with our report here, may be due to a small misalignment in one of the experiments or caused by a limited resolution of the FFT due to the limited field range and asks for the clarification in future experiments.
Another important point for future studies is whether an extra FS exists.

More recently, new quantum oscillation experiments by tunnel diode oscillator (TDO) technique at temperature down to $0.35\,{\rm K}$ was reported~\cite{Bro23}.
\replaced[id=GK]{There,}{The} low frequencies below $1\,{\rm kT}$ with relatively small effective masses, $5.7$--$6.8\,m_0$, are observed at high fields. 
Such frequencies should be easily detected in our dHvA experiments with field-modulation technique.
Further experiments are required to check whether they exist at low fields.

In summary, we detected three fundamental dHvA frequencies for $H\parallel c$-axis with heavy effective masses, corresponding to cylindrical FSs. 
From the anisotropy of the initial slope of $H_{\rm c2}$ and resistivity, we can still consider a possible pocket FS with an extremely large mass.

\section*{Acknowledgements}
We thank Y. \={O}nuki, H. Harima, J. P. Brison, Y. Haga, H. Sakai, S.-i. Fujimori, V. Mineev, Y. Tokunaga, M. Kimata, A. Miyake,  and S. Fujimoto
for fruitful discussion.
This work was supported by KAKENHI (JP19H00646, JP20K20889, JP20H00130, JP20KK0061, JP22H04933), GIMRT (20H0406), ICC-IMR, 
ANR (FRESCO, No. ANR-20-CE30-0020), (NWO), HFML-RU/NWO member of EMFL.


\end{document}